# Imaging topological polar structures in marginally twisted 2D semiconductors

Thi-Hai-Yen Vu[1†], Daniel Bennett[2,3†*], Gayani Nadeera Pallewella[4], Johnathon Maniatis[1], Josh Edwards[5,6], Md Hemayet Uddin[7], Kaijian Xing[1], Pablo Resendiz-Vazquez[1], Seng Huat Lee[8,9], Zhiqiang Mao[8,9], Jack B. Muir[10,11], Linnan Jia[10,11], Jeffrey A. Davis[10,11], Kenji Watanabe[12], Takashi Taniguchi[13], Shaffique Adam[3,14,15], Pankaj Sharma[5,6,16*], Michael S. Fuhrer[1,17,*], Mark T. Edmonds[1,17,18,*]

[1] School of Physics and Astronomy, Monash University, Clayton VIC 3800, Australia

[2] John A. Paulson School of Engineering and Applied Sciences, Harvard University, Cambridge, Massachusetts 02138, USA

[3] School of Electrical and Electronic Engineering, Nanyang Technological University Singapore, 50 Nanyang Avenue, 639798, Singapore

[4] Centre for Advanced 2D Materials, National University of Singapore, 6 Science Drive 2, Singapore 117546

[5] College of Science and Engineering, Flinders University, Adelaide, South Australia 5001 Australia

[6] Flinders Institute for Nanoscale Science and Technology, Flinders University, Adelaide, South Australia, 5042, Australia

[7] Melbourne Centre for Nanofabrication, Victorian Node of the Australian National Fabrication Facility, Clayton 3168, VIC, Australia

[8] Department of Physics, Pennsylvania State University, University Park, PA, 16802, USA

[9] 2D Crystal Consortium, Materials Research Institute, Pennsylvania State University, University Park, PA, 16802, USA

[10] Optical Sciences Centre, Swinburne University of Technology, Hawthorn, 3122, Victoria, Australia

[11] ARC Centre of Excellence in Future Low-Energy Electronics Technologies, Swinburne University of Technology, Hawthorn, 3122, Victoria, Australia

[12] Research Center for Electronic and Optical Materials, National Institute for Materials Science, 1-1 Namiki, Tsukuba 305-0044, Japan

[13] Research Center for Materials Nanoarchitectonics, National Institute for Materials Science, 1-1 Namiki, Tsukuba 305-0044, Japan

[14] Department of Material Science and Engineering, National University of Singapore, 9 Engineering Drive 1, 117575, Singapore

[15] Department of Physics, Washington University in St. Louis, St. Louis, Missouri 63130, United States

[16]ARC Centre of Excellence in Future Low Energy Electronics Technologies, UNSW Sydney, NSW, 2052, Australia

[17] ARC Centre of Excellence in Future Low-Energy Electronics Technology, Monash University, Clayton, 3800, Victoria, Australia

[18] ANFF-VIC Technology Fellow, Melbourne Centre for Nanofabrication, Victorian Node of the Australian National Fabrication Facility, Clayton, VIC 3168, Australia

[†] These authors contributed equally to this work

*Correspondence to: daniel.bennett@ntu.edu.sg, pankaj.sharma@flinders.edu.au michael.fuhrer@monash.edu, mark.edmonds@monash.edu

**Abstract:**

**Moiré superlattices formed in van der Waals heterostructures due to twisting, lattice mismatch and strain present an opportunity for creating novel metamaterials with unique properties not present in the individual layers themselves[1,2]. Ferroelectricity for example, arises due to broken inversion symmetry in twisted and strained bilayers of 2D semiconductors with stacking domains of alternating out-of-plane polarization[3–7]. However, understanding the individual contributions of twist and strain to the formation of topological polar nanostructures remains to be established and has proven to be experimentally challenging. Inversion symmetry breaking has been predicted to give rise to an in-plane component of polarization along the domain walls, leading to the formation of a network of topologically non-trivial merons (half-skyrmions) that are Bloch-type for twisted and Néel-type for strained systems[8]. Here we utilize angle-resolved, high-resolution vector piezoresponse force microscopy (PFM) to spatially resolve polarization components and topological polar nanostructures in marginally twisted bilayer $WSe_2$, and provide experimental evidence for the existence of topologically non-trivial meron/antimeron structures. Our approach can be used to distinguish between Bloch-type and Néel-**

**type merons as well as hybrid structures, allowing us to quantify the separate contributions of strain and twist in a moiré superlattice. This first demonstration of non-trivial real-space topology in a twisted van der Waals heterostructure opens pathways for exploring the connection between twist and topology in engineered nano-devices.**

INTRODUCTION

Stacking two-dimensional (2D) van der Waals (vdW) materials to form heterostructures has the potential to create new physical properties and functionalities. By twisting or straining layers with respect to one another, forming a moiré superlattice, a wide array of emergent properties has been realized such as superconductivity[1], correlated phases[2,9], magnetism[10] and even fractional Chern insulator states[11]. The polar and electromechanical properties of moiré materials have also proven to be remarkably rich, where vdW systems without AB sublattice symmetry become ferroelectric by altering the relative stacking between the layers in order to break centrosymmetry[12,13]. This is demonstrated for transition metal dichalcogenides in Fig. 1(a): where the metals and chalcogens in neighbouring layers are vertically aligned (AB and BA stackings), the mirror plane between the layers is broken, resulting in an out-of-plane polarization via an interlayer electronic charge transfer. The AB and BA stackings have equal and opposite polarization as they are related by a mirror operation. An applied electric field can cause a relative sliding between the layers (vdW sliding) when the potential across the film is larger than the energy barrier between the AB and BA stackings, thus accessing bi-stable ground states, resulting in ferroelectricity[13]. For marginally twisted bilayers, a network of AB and BA stacking domains form, separated by domain walls (DWs) as depicted in Fig. 1(b). For TMDs with small twists, the stacking domains can be identified as a regular network of moiré polar domains (MPDs)[14]. These MPDs can grow and shrink in response to an applied field[15,16] which is mediated by DW bending and motion. Under experimental conditions, due to domain wall pinning, the MPDs deform irreversibly, resulting in a switchable remnant polarization when the electric field is applied and removed[3–5,7].

Recently, it was proposed that the MPDs also have a spatially varying in-plane polarization component[8,17]. Although the DW stacking, where locally there is a relative shift of half a unit cell diagonal between the layers, does not possess a mirror symmetry, it is invariant under a mirror operation plus a non-symmorphic translation of half a unit cell diagonal, preventing any out-of-plane polarization. An in-plane polarization is not prevented by any symmetry however, and the DW stacking has an in-plane polarization which is parallel to the relative shift between the layers. Thus, the MPDs in twisted and strained bilayers exhibit winding, transitioning from in-plane along the DWs to out-of-plane exactly at the AB/BA domain centers. In homobilayers, the winding in each MPD, including both the domain interiors and domain walls, is topologically nontrivial, with winding numbers of ±½, and the domain structure forms a regular network of merons and antimerons (half-skyrmions and half-antiskyrmions)[8]. For twisted bilayers, the in-plane polarization circulates around the AB/BA domain centers, and the merons are of Bloch type, whereas for strained bilayers, i.e. one with a small lattice mismatch between the layers, the in-plane polarization flows into and out of the domain centers, and the merons are of Néel type. The meron topological texture that exhibits out-of-plane vectors at the core region and gradually changes to in-plane vectors has been found in both ferromagnetic and ferroelectric materials[18–20]. Topological magnetic structures have applications in high density data storage due to their stability, as well as in logic gates[21]. Topological polar structures, typically observed in oxide perovskite nanostructures[22], are thought to be advantageous for ultrafast energy storage, with phonon frequencies typically in THz range[23].

While very appealing both in terms of fundamental physics in terms of investigating the connection between twist and topology and potential applications in nano-engineered devices that enable on-demand creation and manipulation of polar topological objects, the topological nature of the MPDs has been difficult to experimentally verify, primarily due to two factors. First, as the polarization in vdW materials is purely electronic, the in-plane component cannot be measured using the same techniques which are employed to measure the out-of-plane polarization, such as the Kelvin probe force microscopy[13], resistance measurements[4] and electron microscopy[7]. The in-plane polarization could in principle be measured from the lateral deflection from a piezoresponse force microscopy (PFM) tip, although the second issue is that in systems with small twist angles (large moiré periods), typically 1-2 degrees, significant lattice relaxation occurs[24,25], leading to sharp domain structures, with the in-plane polarization confined to the narrow domain walls, with widths of order 1nm. Resolving the in-plane polarization would thus either require a very fine resolution, or larger domain walls, such as those in systems with very small twist angles (very large moiré periods), < 0.5°. A nonzero phase has been measured along the domain walls in bilayer systems[26], primarily graphene, which is ordinarily nonpolar, although this was attributed to flexoelectricity due to the large strain gradient across narrow walls.

Here we report the observation of both out-of-plane and in-plane polarization in four marginally twisted $WSe_2$ bilayer samples all with twist angle around 0.1° using PFM measurements at room temperature. By performing angle-resolved PFM measurements, we directly resolve the in-plane polarization, and we show that each MPD exhibits a clear winding confirming the existence of a topologically nontrivial meron-antimeron network in twisted bilayers. This is in excellent agreement with density functional theory (DFT) and molecular dynamics (MD) calculations predicting in-plane polarization is narrowly confined to the domain walls separating the polar domains. We identify a network of Bloch-type merons and antimerons in twisted bilayers, characterized by polarization parallel to domain walls. In regions with twist and strain, we detect hybrid merons and antimerons, exhibiting both Bloch and Néel characteristics, with polarization components parallel and perpendicular to domain walls. Our technique differentiates between heterostructures formed by twist versus lattice mismatch and can quantify the contributions of each in a moiré superlattice, which is not possible with the out-of-plane polarization measurements alone.

RESULTS

We performed non-invasive high-resolution piezoresponse force microscopy measurements to investigate the polar nanostructures and properties of marginally twisted $WSe_2$. As shown optically in Fig. S1(a) and schematically in Fig. 1(c) we fabricated four separate nearly 0° twisted bilayer $WSe_2$ devices using the dry transfer technique (for details on sample fabrication, see Methods). Using second harmonic generation (see Fig. S1(b)) the twist angle near 0° was confirmed prior to PFM imaging. The exact twist angle in the four twisted bilayer $WSe_2$ devices ranging from 0.03 to 0.13° was determined using the moiré periodicity extracted directly from spatially resolved PFM measurements. Figure 1(d) shows a large area (2.2 μm × 2.2 μm) vertical PFM phase image taken in the twisted bilayer region of one of the $WSe_2$ devices. In agreement with previous experimental studies[5,7], the vertical PFM phase map, which is related to the magnitude of the out-of-plane polarization and thus the piezoelectric coefficient (Fig. 1(d)) shows a clear triangular pattern, with AB and BA domains having opposite piezoelectric contrast. Evidence of the opposite out-of-plane polarizations which are relatively uniform across the domains and consistent with DFT and MD calculations shown in Fig. S2(f-g) and previous results[5,7]. However, the periodicity of the triangular patterns varies as the local twist angle and/or layer-dependent strain changes due to wrinkles and bubbles that are observed in the corresponding AFM topography image of the same region (Fig. S2(a-b)). The enhanced vertical PFM signal near the DWs (as shown in Fig. S2(e)) is due to large strain gradients which result in a polar response through flexoelectricity[26], which couples to piezoelectric stresses, resulting in two strong lines at the DWs in the vertical PFM signals. The in-plane flexoelectric response however is negligible[26], see Section S3 for discussion.

By capturing vertical and lateral deflections of the reflected laser on the quadrant photodiode detector independently, PFM can be used to detect not just the out-of-plane but also in-plane polarization responses. Lateral deflection results from in-plane torsion perpendicular to the cantilever axis. On the other hand, vertical deflection includes surface deformation resulting from both out-of-plane polarization and in-plane polarization components parallel to the cantilever axis (See Fig. S2(c)). By performing lateral PFM amplitude and phase measurements in bilayer regions with twist angle (~0.13°) we observe that the lateral response in Fig. 1(e) is confined solely to the narrow domain walls, with minimal response observed in the interior of AB and BA domains. This distinctly suggests the presence of in-plane polarization localized to the domain walls. The corresponding lateral phase image in Fig. 1(f) which indicates the polarization direction, shows there is a clear 180º phase switch between the alternating domain walls (histogram analysis of the phase switch is found in SI Fig S3).

The theoretically calculated in-plane polarization $P_{\parallel}(r)$ in Fig. 2(a)-(b) shows the direction of the in-plane polarization field in the domain walls is different for strained bilayers, i.e. one with a small lattice mismatch between the layers, and for twisted bilayers, i.e. two layers twisted with respect to each other. For twisted bilayers in Fig. 2(a), the in-plane polarization is parallel to the domain walls, pointing towards the narrow AA regions, whilst for strained bilayers in Fig. 2(b) the polarization is perpendicular to the domain walls, pointing towards (away from) the AB (BA) domain centers. To confirm these predictions, we use the lateral deflection to observe whether the phase of the PFM signal changes depending on the domain wall orientation. The illustration in Fig. 2(c) explains how in-plane polarization ($\vec{P}$) affects sample deformation and thus, affects lateral deflection during a PFM measurement. A single lateral PFM image allows determination of the component of in-plane polarization perpendicular to the tip ($P_x$) and the polarization sign as shown in the bottom panel of Fig. 2(c). To fully map the in-plane polarization vector field, we utilize a well-established angle resolved vector PFM technique[27], where PFM measurements taken at different angles between sample and cantilever axis are added together to construct a vector map of the in-plane polarization.

Figures 2(d) and 2(e) show representative in-plane PFM measurements on a twisted $WSe_2$ device with twist angle of 0.13°, at two different orientations of the cantilever axis relative to the sample, as illustrated above each panel. The PFM phase along the domain walls changes as a function of the angle and allows an unambiguous assignment of the polarisation directions along the wall, which closes around the triangular domains in a network of clockwise and anticlockwise fashion. Interestingly, this region is close to a bubble, where the MPDs may be affected by both twist and strain due to the bubble. Consequently, a small but measurable phase is detected for the walls that are aligned with the cantilever axis (i.e. Fig. 2d, domain wall at the bottom; Fig. 2e, domain wall at the right), suggesting the presence of a simultaneous polarization component perpendicular to these walls. Figure 2(f) shows the result of the vector PFM analysis constructed from Figs. 2(d) and (e) as well as several other images (for further details on this method, see SI section S4). These in-plane vector PFM maps visualize and make the winding of the polarization immediately clear. In this region the shape of the elucidated in-plane polarization field runs slightly diagonal along the domain walls, demonstrating this moiré superlattice results from both twist (Fig. 2a) and strain (Fig. 2b) and establishes in-plane/torsional PFM as an effective and direct method for differentiating between twisted and strained moiré superlattices.

To gain statistically robust information, full angular dependent PFM measurements were performed on a set of three marginally twisted $WSe_2$ devices with twist angles of 0.03º, 0.04º and 0.05º, whose in-plane vector PFM maps are shown in Fig. 3(a)-(c) respectively. The corresponding angular dependent lateral PFM amplitude and phase images are shown in SI Section S5. In these regions (which were scanned away from bubbles and wrinkles) the in-plane polarization is almost entirely parallel to the domain walls, meaning that within these regions of bilayer $WSe_2$ the MPD arises entirely due to twist. The angular dependence of the polarization response was further confirmed from lateral PFM measurements as a function of the relative angle (φ) between each domain wall orientation and the cantilever direction. For each domain, the polarization response was quantified by measuring the PFM lateral amplitude after performing a background subtraction and combining it with the sign of the polarization direction determined from the lateral phase. The resulting data points were normalized then plotted as a function of the relative angle φ, in Fig. S5-S7 and then fitted using the function: $y = y_0 + A sin\left(\pi \frac{\varphi - \varphi_c}{w}\right)$ where $\varphi_c$ is a phase shift, $w$ is the period, $A$ is the amplitude, and $y_0$ is the offset. This yielded phase shifts of -1.7°, 2.4° and 0.6° for the 0.03°, 0.04° and 0.05° twisted bilayer $WSe_2$ devices respectively, confirming the in-plane polarization is almost entirely parallel to the domain walls. This is in excellent agreement with theoretical predictions (Fig. 2(a) and Fig. 4(a),(b)) that, after lattice relaxation, the in-plane polarization is confined to the domain walls, as the interior of the domains has nearly uniform AB/BA stacking, and in-plane polarization is prevented by $C_3$ symmetry. We note that for these measurements, the in-plane phase difference was much smaller than the ideal 180° but is still sufficient to demonstrate the topological winding (see Section S6 of SI). The observed much smaller phase difference can arise from a range of factors, including losses associated with tip-sample mechanical coupling, contact stiffness, and other non-idealities, as has been reported in many of the previously published works, including on classical ferroelectric oxides, where the reported phase difference ranges from a few degrees to tens of degrees[26,28–30]. Moreover, our measurements were performed across various PFM tips and scanning conditions (e.g. PFM excitation frequency and amplitudes, scan angles etc.) refer to Section S8 and S9 of the SI, and topological winding of the in-plane polarization vector field is *always* observed in moiré superlattices, which further attests to the intrinsic nature of the topological winding.

First-principles DFT calculations were performed in order to validate the observation of in-plane polarization in the twisted and strained regions of bilayer $WSe_2$ (see Methods), and to verify the topological nature of the MPDs. The out-of-plane and in-plane polarization in bilayer $WSe_2$ was calculated as a function of relative stacking between the layers. The out-of-plane polarization has a maximum and minimum at the AB and BA stackings, and is zero for the AA and DW stackings. The in-plane polarization is zero for the AA, AB and BA stackings, and is largest for the DW stacking. In order to accurately describe the polarization field observed in the sample, the significant lattice relaxation which occurs at small angles[23] is taken into account (see Methods). The resulting in-plane polarization for bilayer $WSe_2$ with a relative twist angle of 0.13° is shown in Fig. 4(a). The in-plane polarization is confined to and points parallel to the narrow domain walls, with negligible polarization inside the MPDs, in excellent agreement with the vector maps in Fig. 2(f) and 3(a)-(c). Combining the in-plane and out-of-plane (Fig.S2), the winding (topological charge) of the total polarization was calculated (Fig.4(c)). We note that total polarization is calculated as the dipole moment divided by the volume $V = Ad$, where $A$ is the in-plane area of the primitive unit cell, and the interlayer separation $d$ is taken to be the separation between the metals in each layer. The winding is of opposite sign in the AB and BA MPDs, and integrates to ±½, confirming the topological nature of the experimentally measured polarization in marginally twisted bilayer $WSe_2$: a network of Bloch type merons and antimerons (Fig.4(e)).

In addition to DFT calculations, large-scale MD calculations were also performed in order to determine the structure of bilayer $WSe_2$ twisted at an angle of 0.13° (see Methods). This numerical calculation offers the advantage of encompassing all atoms within the moiré supercell. Starting with two layers of $WSe_2$ with a global twist of 0.13°, the system was relaxed in order to determine the equilibrium geometry of the bilayer. The resulting in-plane polarization is shown in Fig. 4(b), which also shows polarization confined to the domain walls and circulating around the domain centers. Combining the in-plane and out-of-plane (Fig.S2), the topological charge was calculated (Fig.4(d)), which also indicates a network of Bloch type merons and antimerons. We note that, although the structural properties differ between DFT calculations, which predict sharper stacking domains, and more realistic large-scale MD calculations, both methods verify the nature of the in-plane polarization and out-of-plane polarization, as experimentally measured, and confirm the topological nature of the MPDs.

## DISCUSSION

In this work, we demonstrate that in-plane PFM measurements can be used to detect complex polarization textures in moiré superlattices. We have directly confirmed that the MPDs in a twisted TMD have an in-plane texture, and this can be used to directly probe whether MPDs arise due to twist or strain, thus providing a simple and effective method to differentiate between the two components in moiré superlattices. Furthermore, by using angular-dependent in-plane PFM we demonstrate the in-plane polarization in twisted $WSe_2$ is parallel to the domain walls, and winds around the AB and BA domain centers, in excellent agreement with two independent theoretical predictions. This confirms the topological nature of the MPDs, i.e. that they form a regular network of merons and antimerons. In contrast to the polar skyrmions typically observed in oxide perovskite superlattices[31] and recently in moiré oxide heterostructures[32], typically tens of nanometers in thickness, the meron-antimeron network we discover is the first such polar topological structure observed in a truly two-dimensional system (approximately 1 nm thick).

The techniques developed in this work represent a novel method for measuring complex and topological polar structures, which, for the case of bilayers comprised of TMDs or hBN, are difficult or impossible to determine using other methods typically used to detect topological polar structures in oxide perovskites[19,22,23,31], as the polarization is electronic and are difficult to determine solely by measuring the individual displacements of the atoms. We anticipate these techniques will play an important role in further exploration and understanding of the connections between twist and topology in other twisted 2D systems, across a wide span of twist angles and layer configurations. Finally, non-trivial real-space topology in twisted van der Waals heterostructures will open pathways in engineered nano-devices, where the manipulation of polar topology can be realised via electric gating[8], mechanical deformation[33] or by engineering substrates and/or additional 2D layers with tailored strain, doping or twisting.

**Acknowledgements:**

We would like to acknowledge C. Dreyer for valuable discussions, and GP and SA acknowledge helpful discussions with Mohammed Al-Ezzi and Christophe De Beule.

**Funding:**

M. T. E, M. S. F., E. V. Y, S. A., K. X. acknowledge funding support from ARC Discovery Project DP200101345. M. T. E. acknowledges funding support from ARC Future Fellowship FT220100290. P. S. acknowledges funding support from ARC Discovery Project DP240102137. M. T. E., M. S. F., J. B. M·, L. J., J. A. D. acknowledge funding support from ARC Centre for Future Low Energy Electronics Technologies (FLEET) CE170100039. This work was performed in part at the Melbourne Centre for Nanofabrication (MCN) in the Victorian Node of the Australian National Fabrication Facility (ANFF). D.B. acknowledges support from the US Army Research Office (ARO) MURI project under grant No. W911NF-21-0147, support from the Nanyang Technological University Startup Grant (Award Number 025661-00003) and from the Simons Foundation award No. 896626. GP and SA acknowledge support from the Singapore National Research Foundation Investigator Award (NRF- NRFI06-2020-0003). K.W. and T.T. acknowledge support from the JSPS KAKENHI (Grant Numbers 21H05233 and 23H02052) and World Premier International Research Center Initiative (WPI), MEXT, Japan. Support for crystal growth and characterization was provided by the National Science Foundation through the Penn State 2D Crystal Consortium-Materials Innovation Platform (2DCC-MIP) under NSF cooperative agreements DMR-2039351.

**Author contributions:**

M.T.E conceived the project. T.-H.-Y.V designed the experiments, fabricated the device and performed the PFM measurements. K.X assisted with sample fabrication. Support for the PFM measurements was provided by P.S and H.U. Second Harmonic Generation was provided by J.B.M, L.J, and J.A.D. D.B, G.N.P and S.A performed the theoretical calculations. S.H.L and Z.M grew the $WSe_2$ and K.W. and T.T grew the hBN crystal. T.-H.-Y.V analyzed the experimental data. M.T.E, M.S.F supervised the project. T.-H.-Y.V, D.B, M.S.F, and M.T.E wrote the manuscript, with input from all other authors.

**Competing interests:**

The authors declare they have no competing interest.

**Data and Materials Availability:**

All data needed to evaluate and reproduce the results in the paper are present in the paper and/or the Supplementary Materials

**Note Added:**

After the submission of our manuscript, a number of other works were published which explored various aspects of in-plane polarization in twisted bilayers[34–38].

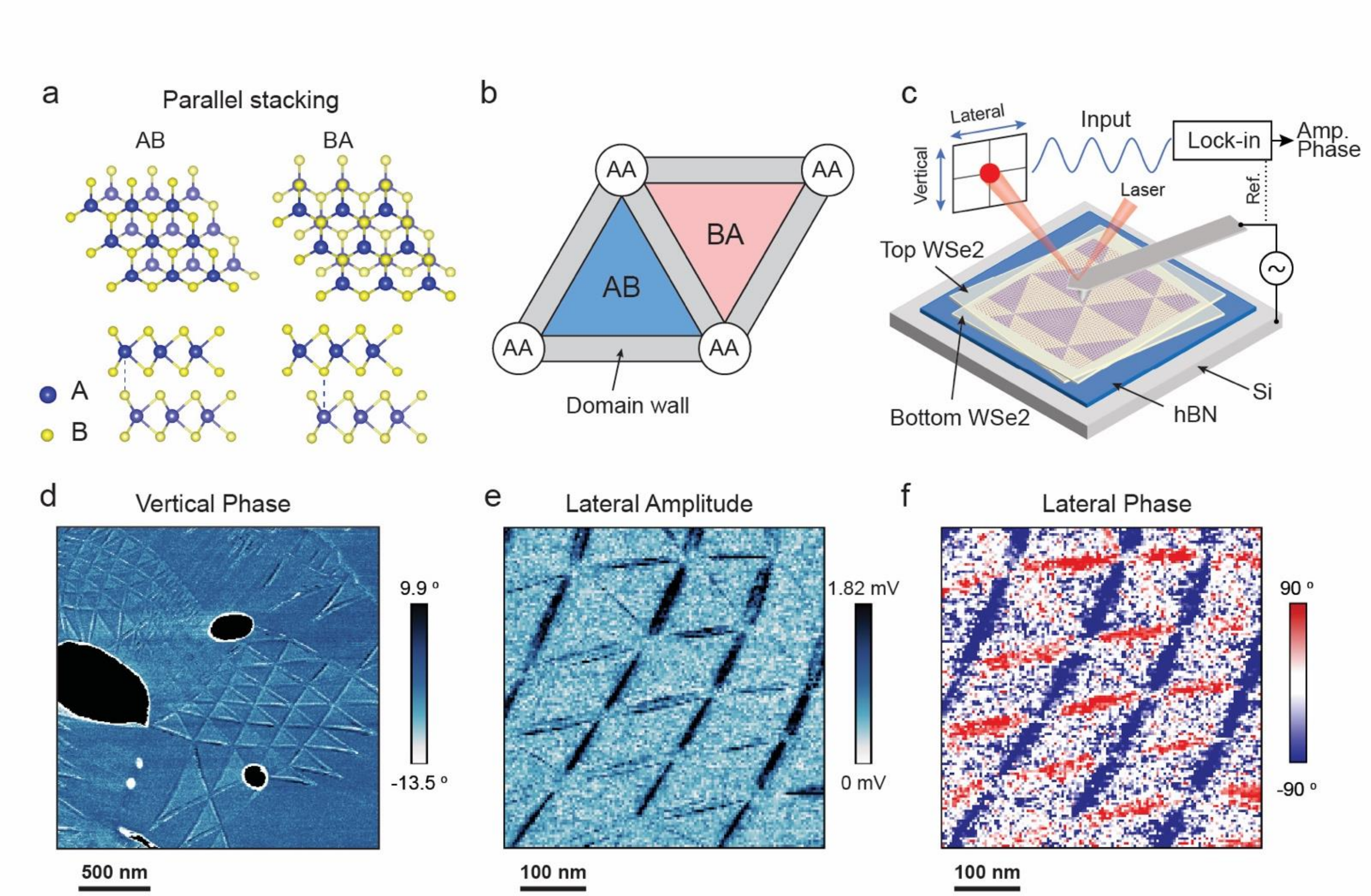

**Figure 1. Piezoresponse force microscopy visualization of marginally twisted bilayer TMDs. a,** Illustration of bilayer $AB_2$ in the parallel stacking, where A is a transition metal (Mo, W) and B is a chalcogen (S, Se). The energetically favourable AB and BA stackings are shown, where the A and B atoms in neighbouring layers are vertically aligned. **b,** Schematic depicting a moiré superlattice with AB (blue), BA (red), AA (white) stacking regions and domain walls (grey). **c,** Schematic of PFM on a moiré pattern formed by marginally twisted bilayer $WSe_2$ **d,** Large-area (2.2 μm × 2.2 μm) vertical phase PFM image. **e, f,** Lateral amplitude and phase PFM measurements taken within the same twisted bilayer region, respectively.

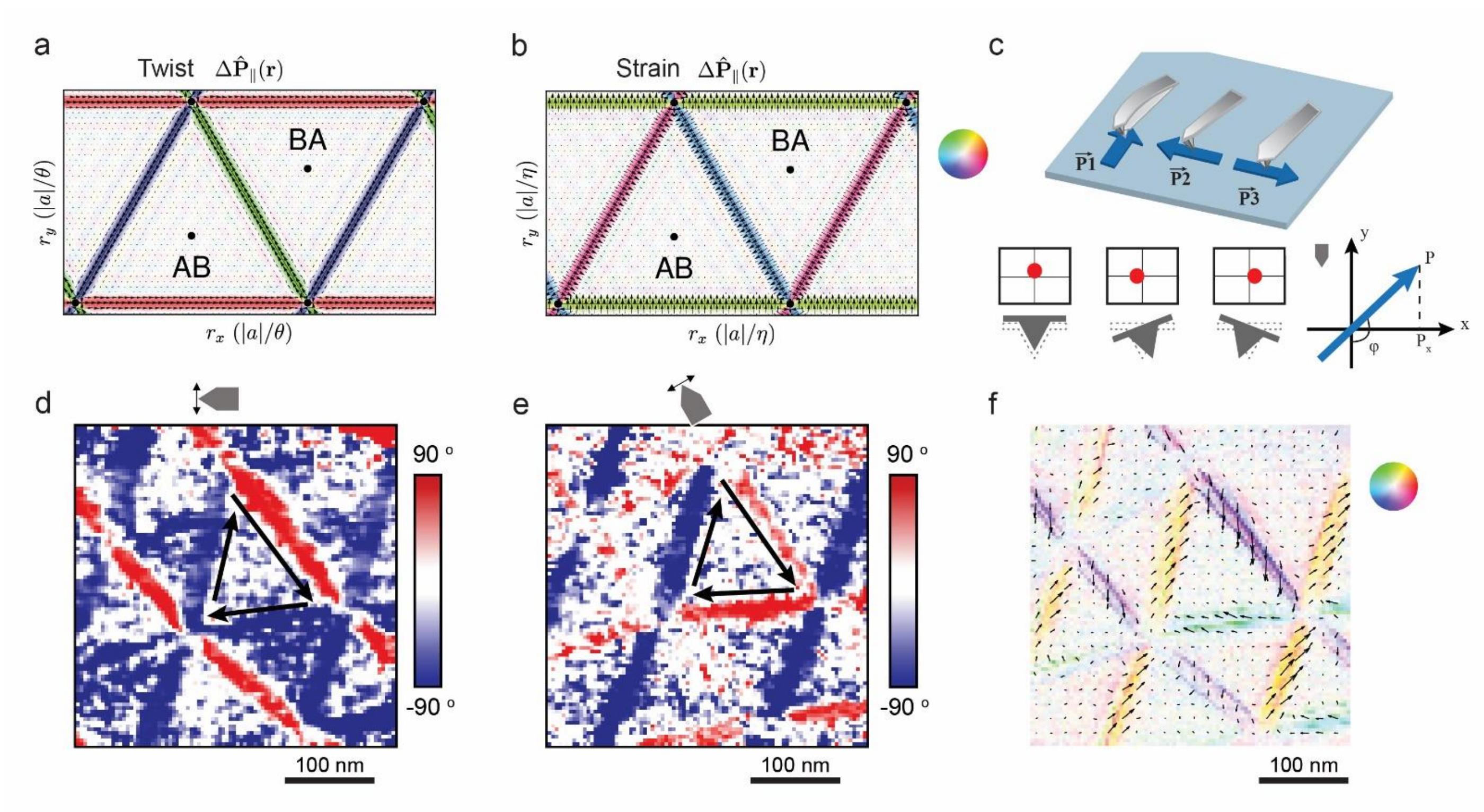

**Figure 2. In-plane polarization distribution in marginally twisted bilayer $WSe_2$.** Theoretically calculated changes in in-plane polarization $P_{\parallel}(r)$ for **a** twisted and **b** biaxially strained (lattice mismatch) $WSe_2$ after lattice relaxation. The lattice is scaled by the twist angle θ and lattice mismatch η, respectively. **c,** The top panel illustrates how in-plane polarization vector $\vec{P}$ (blue arrow) affects sample deformation and cantilever lateral torsion during a PFM measurement. The bottom panel shows deflection on the quadrant photodiode detector of $\vec{P}_1$(left), $\vec{P}_2$ (middle), $\vec{P}_3$ (right) denoted in the top panel and the intensity of polarization ($P_x$) as a function of the in-plane polarization at a certain angle (φ) between the cantilever (grey arrow) and the polarization vector (blue arrow). **d, e,** Phase images of lateral PFM measurements, performed on a 0.13° region of bilayer $WSe_2$. The grey insets represent the orientation of the cantilever and the double headed arrows represent the scan direction. The one-headed arrows indicate the direction of the in-plane polarization within the domain walls. The direction was determined based on the convention in the bottom panel of **c**. **f**. Angle-Resolved-PFM (AR-PFM) colorized vector field map generated from lateral PFM images in **d-e**.

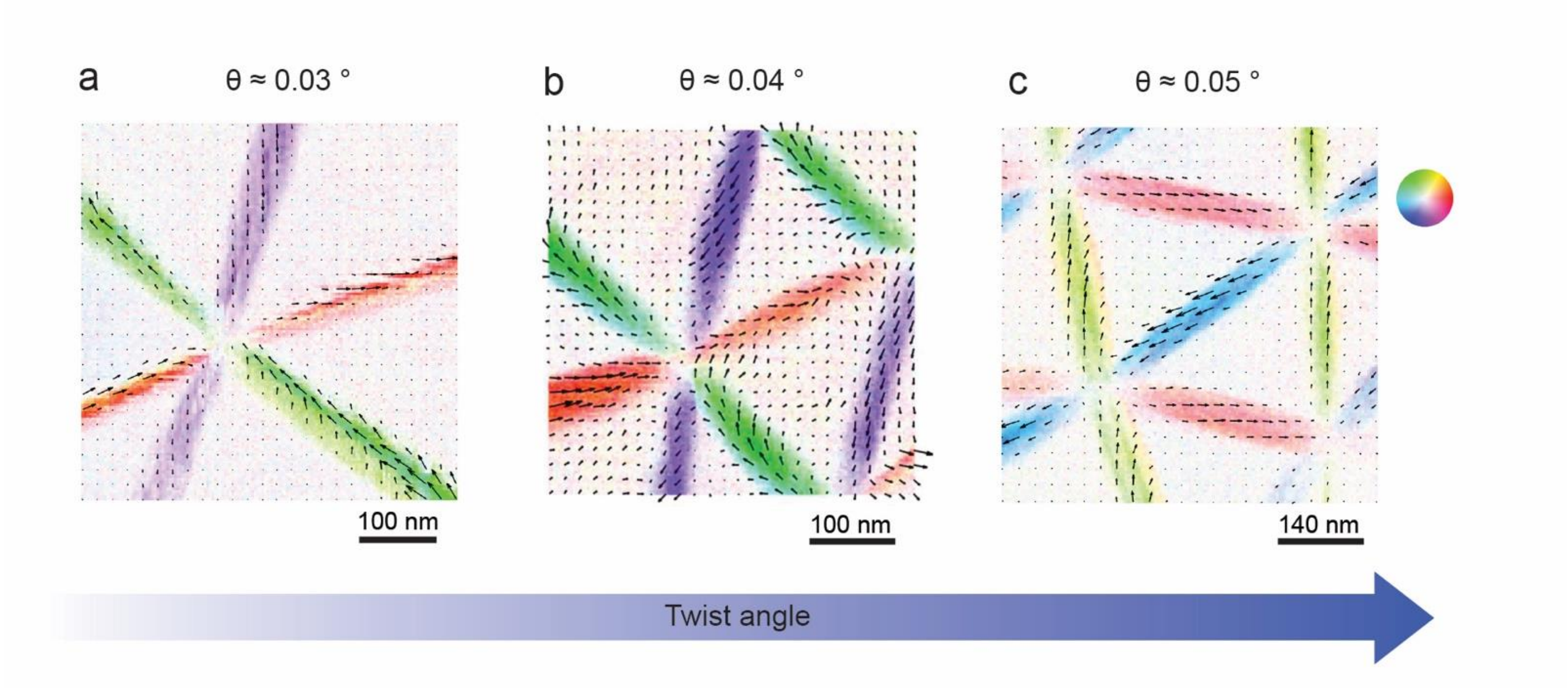

**Figure 3. Imaging the angular dependence of in-plane polarization. a-c**, Angle-Resolved-PFM (AR-PFM) colorized vector field maps with varying twist angles generated from lateral PFM images (see SI Section S5).

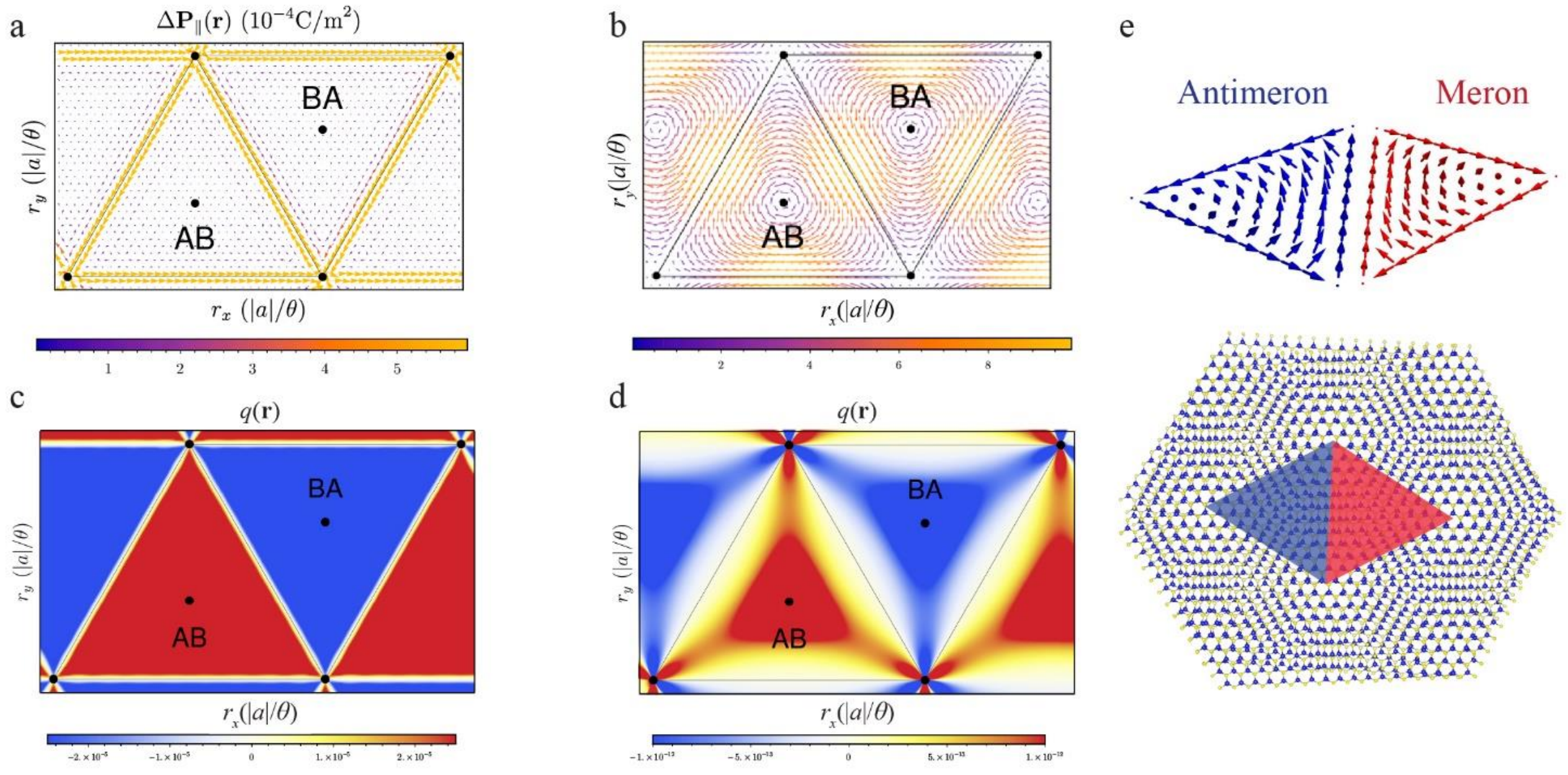

**Figure 4. Theoretical calculation of polarization textures in twisted $WSe_2$. a,** In-plane polarization in a 0.13° twisted bilayer $WSe_2$, obtained from DFT calculations. The effects of lattice relaxation on the stacking domains are included. **b,** In-plane polarization in a 0.13° bilayer $WSe_2$, obtained from MD simulations. In-plane polarization is acquired through parameterization facilitated by DFT calculations**. c, d,** Topological charge, i.e. the winding of the total polarization field, from **c** DFT calculations and **d** MD simulations. The polarization is normalized everywhere except for the AA stacking, where both out-of-plane and in-plane components are zero. **e,** Schematic illustration of a twisted bilayer where the two AB and BA polar domains are highlighted, and the merons and antimerons which form are sketched above.

**MATERIALS & METHODS:**

Sample fabrication

We used blue tape to exfoliate hexagonal boron nitride (*h*BN) onto a conductive silicon wafer, then selected a flat flake with suitable thickness under the optical microscope. The $WSe_2$ monolayers were prepared on polydimethylsiloxane (PDMS) substrates using the same method. The monolayer nature was confirmed by both optical contrast and photoluminescence spectroscopy. Their orientation was determined by second harmonic generation spectroscopy. The $WSe_2$ monolayers were transferred onto the *h*BN one by one by dry transfer method and aligned to each other to obtain marginally twisted bilayer $WSe_2$. After each transfer, the sample was cleaned by diisopropylamine and annealed in an Argon environment at 100° C for 1 hour. Finally, the sample was annealed in ultra-high vacuum at 270°C for 8 hours before measurements.

Piezoresponse force microscopy

PFM measurements were performed on a commercial scanning probe microscope, i.e. Bruker Dimension Icon, at room temperature with a Nanoscope 6 Controller. We used conductive Platinum-Iridium coated Bruker probes with tip end radius of 25 nm and spring constant of around 3 N $m^{-1}$. AC bias was applied through the tip, and induced sample deformation whose amplitude and phase represent the magnitude of the piezoelectric coefficient and the polarization direction of the response, was detected respectively. For both vertical and lateral PFM, AC bias voltages were set in the range of 2V – 3V and frequencies were around 250-280 kHz. In our measurement, the tip frequency was set near the contact resonance frequency. Contact resonance frequency is where the tip and the sample are in resonance as shown in Fig. S9. Here, we have a very good signal-to-noise ratio but mechanical response and piezoelectric response will be mixed. To avoid these issues measurements for all samples (i.e. twist angle of 0.03°, 0.04º, 0.05° and 0.13°) were performed at totally off-resonance frequencies (see Section S8 of SI).

First-principles calculations

First-principles density functional theory (DFT) calculations were performed using the SIESTA[39] code, using PSML norm-conserving pseudopotentials, obtained from Pseudo-Dojo[40]. SIESTA employs a basis set of numerical atomic orbitals (NAOs), and double-ζ polarized (DZP) orbitals were used for all calculations. A Monkhorst-Pack **k**-point grid[41] of 12 × 12 × 1 was used for the initial geometry relaxations, and a mesh of 18x18x1 was used to calculate the polarization. Calculations were converged until the relative changes in the hamiltonian and density matrix were both less than $10^{-6}$. The C09[42] van der Waals exchange-correlation functional was used to treat the long-range interactions between the layers. A dipole correction was employed in the vacuum region to prevent dipole-dipole interactions between periodic images.

The top layer was translated along the unit cell diagonal over the bottom layer, which was held fixed. At each point a geometry relaxation was performed to obtain the equilibrium layer separation, while keeping the in-plane lattice vectors fixed. The out-of-plane and in-plane polarization were then obtained by calculating the Berry phases of the Bloch states. The data were fitted to Fourier expansions which respect the $\mathcal{C}_3$ rotation symmetry t-$WSe_2$[8].

Lattice Relaxation

Lattice relaxation calculations were performed following the methodology in Refs.[15,16], for t-WSe2 twisted at an angle of $\theta$=0.13° with respect to the ideal rhombohedral stacking (perfectly aligned layers). The total energy of t-$WSe_2$ is given by

$$V_{\mathrm{tot}} = \int \mathcal{V}_{\mathrm{tot}}(\mathbf{x}+\mathbf{u}(\mathbf{x}))d\mathbf{x}$$

$$\mathcal{V}_{\mathrm{tot}}(\mathbf{x}) = \mathcal{V}_{\mathrm{stack}}(\mathbf{x})+\mathcal{V}_{\mathrm{elastic}}(\mathbf{x})$$

where $\mathcal{V}_{\mathrm{tot}}(\mathbf{x})$ is the total energy density as a function of relative stacking $\mathbf{x}$ between the layers, and $\mathbf{u(x)}$ is a displacement field which describes the relaxation of the bilayer from its rigid twisted configuration. The integration is performed in "configuration space"[24], in terms of the relative stackings between the layers, all of which are contained in a single primitive cell of $WSe_2$. The total energy density is given as a sum of two independent terms. The stacking energy, $\mathcal{V}_{\mathrm{stack}}(\mathbf{x})$,

$$\mathcal{V}_{\mathrm{stack}}(\mathbf{x})=\sum_n \mathcal{V}^e_n\, \phi^e_n(\mathbf{x})+\mathcal{V}^o_n\, \phi^o_n(\mathbf{x}),$$

describes the vdW or cohesive energy between the layers. It is written as a Fourier expansion using even and odd $\mathcal{C}_3$ symmetric basis functions $\phi^{e/o}_n$:

$$\phi^e_1 = \cos(2\pi x) + \cos(2\pi y) + \cos(2\pi(x+y))$$

$$\phi^e_2 = \cos(2\pi(x-y)) + \cos(2\pi(2x+y)) + \cos(2\pi(x+2y))$$

$$\phi^e_3 = \cos(4\pi x) + \cos(4\pi y) + \cos(4\pi(x+y))$$

$$\phi^o_1 = \sin(2\pi x) + \sin(2\pi y) - \sin(2\pi(x+y))$$

$$\phi^o_2 = \sin(2\pi(y-x)) + \sin(2\pi(2x+y)) - \sin(2\pi(x+2y))$$

$$\phi^o_3 = \sin(4\pi x) + \sin(4\pi y) - \sin(4\pi(x+y))$$

where x and y are fractions of the primitive lattice vectors of $WSe_2$. The elastic energy,

$$\boldsymbol{\mathcal{V}}_{\text{elastic}}(\mathbf{x}) = \frac{\theta^2}{2}\left[B(\partial_x u_y - \partial_y u_x)^2 + \mu((\partial_x u_y + \partial_y u_x)^2 + ((\partial_x u_x - \partial_y u_y)^2)\right],$$

describes the elastic penalty of deforming the layers where $B$and $\mu$ are the bulk and shear modulus respectively. The total energy $V_{\text{tot}}$ was minimized to obtain the displacement field $\mathbf{u}(\mathbf{x})$ for fixed values of $\theta$.

The out-of-plane polarization, which is odd with respect to stacking, is given by

$$P_\perp(\mathbf{x}) = \sum_i P^\perp{}_n\, \phi^o{}_n(\mathbf{x}).$$

The in-plane polarization is even with respect to stacking, and thus the vector basis functions can be given by $\nabla_x \phi^o{}_n(\mathbf{x})$:

$$\mathbf{P}_{\parallel}(\mathbf{x}) = \sum_i P^{\parallel}{}_n\, \nabla_x \phi^o{}_n(\mathbf{x}).$$

The coefficients $P^\perp{}_n$ and $P^{\parallel}{}_n$ are obtained by fitting the polarization obtained from DFT calculations to $\boldsymbol{C}_3$-symmetric odd and even scalar fields and vector fields, respectively.

The resulting polarization field including the effects of lattice relaxation $\mathbf{P}(\mathbf{x}+\mathbf{u}(\mathbf{x}))$ for $WSe_2$ with a twist angle of 0.13° is shown in Figs. 4 (a) and (b).

The winding of the polarization

$$\mathbf{q}(\mathbf{x}) = \mathbf{P}(\mathbf{x})\cdot(\partial_x \mathbf{P}(\mathbf{x}) \times \partial_y \mathbf{P}(\mathbf{x})),$$

where **x**=(x,y) and **P**(**x**) is normalized, is shown in Fig. 4(c). The winding was calculated following the methodology in Ref.[16], on a fine real space grid, offset from the AA stacking by half a grid spacing, the only point where the normalized polarization is not well-defined. Integrating the winding in the AB and BA domains yields a total winding of $Q_{AB}$=+½ (meron) and $Q_{BA}$=-½ (meron), respectively.

Molecular Dynamics calculations

In addition to DFT calculations, lattice relaxations were also performed using molecular dynamics (MD) simulations. We utilized MD simulations employing the Large-scale Atomic Molecular Massively Parallel Simulator (LAMMPS)[43] and classical interatomic force field models for atomic relaxation.

For the case of twisted $WSe_2$, we applied the KC potential for interlayer interactions and the SW potential for intralayer intralayer interactions with SW/mod style[44,45]. Lattice relaxation calculations were conducted for a commensurate twist angle $\theta = 0.13°$, involving approximately 1 million atoms. Utilizing the relaxed atomic positions, the in-plane displacement is computed between the bottom and top layers. Subsequently, the determination of the out-of-plane and in-plane of polarization, as well as the topological charge is achieved via parameterization, facilitated by the DFT calculations. The quantitative charge values differ between MD and DFT calculations due to the utilization of distinct grid sizes in the analysis. Nevertheless, qualitatively, the AB and BA configurations exhibit opposite winding behaviour and converge to ±½.

# Supplementary Information
# Imaging topological polar structures in marginally twisted 2D semiconductors

Thi-Hai-Yen Vu[1], Daniel Bennett[2,3*], Gayani Nadeera Pallewella[4], Johnathon Maniatis[1], Josh Edwards[5,6], Md Hemayet Uddin[7], Kaijian Xing[1], Pablo Resendiz-Vazquez[1], Seng Huat Lee[8,9], Zhiqiang Mao[8,9], Jack B. Muir[10,11], Linnan Jia[10,11], Jeffrey A. Davis[10,11], Kenji Watanabe[12], Takashi Taniguchi[13], Shaffique Adam[3,14,15], Pankaj Sharma[5,6,16*], Michael S. Fuhrer[1,17,*], Mark T. Edmonds[1,17,18,*]

[1] School of Physics and Astronomy, Monash University, Clayton VIC 3800, Australia

[2] John A. Paulson School of Engineering and Applied Sciences, Harvard University, Cambridge, Massachusetts 02138, USA

[3] School of Electrical and Electronic Engineering, Nanyang Technological University Singapore, 50 Nanyang Avenue, 639798, Singapore

[4] Centre for Advanced 2D Materials, National University of Singapore, 6 Science Drive 2, Singapore 117546

[5] College of Science and Engineering, Flinders University, Adelaide, South Australia 5001 Australia

[6] Flinders Institute for Nanoscale Science and Technology, Flinders University, Adelaide, South Australia, 5042, Australia

[7] Melbourne Centre for Nanofabrication, Victorian Node of the Australian National Fabrication Facility, Clayton 3168, VIC, Australia

[8] Department of Physics, Pennsylvania State University, University Park, PA, 16802, USA

[9] 2D Crystal Consortium, Materials Research Institute, Pennsylvania State University, University Park, PA, 16802, USA

[10] Optical Sciences Centre, Swinburne University of Technology, Hawthorn, 3122, Victoria, Australia

[11] ARC Centre of Excellence in Future Low-Energy Electronics Technologies, Swinburne University of Technology, Hawthorn, 3122, Victoria, Australia

[12] Research Center for Electronic and Optical Materials, National Institute for Materials Science, 1-1 Namiki, Tsukuba 305-0044, Japan

[13] Research Center for Materials Nanoarchitectonics, National Institute for Materials Science, 1-1 Namiki, Tsukuba 305-0044, Japan

[14] Department of Material Science and Engineering, National University of Singapore, 9 Engineering Drive 1, 117575, Singapore

[15] Department of Physics, Washington University in St. Louis, St. Louis, Missouri 63130, United States

[16]ARC Centre of Excellence in Future Low Energy Electronics Technologies, UNSW Sydney, NSW, 2052, Australia

[17] ARC Centre of Excellence in Future Low-Energy Electronics Technology, Monash University, Clayton, 3800, Victoria, Australia

[18] ANFF-VIC Technology Fellow, Melbourne Centre for Nanofabrication, Victorian Node of the Australian National Fabrication Facility, Clayton, VIC 3168, Australia

[†] These authors contributed equally to this work

*Correspondence to: dbennett@seas.harvard.edu, pankaj.sharma@flinders.edu.au

michael.fuhrer@monash.edu, mark.edmonds@monash.edu

**Table of Contents**

## 1. Optical image and second harmonic generation of twisted bilayer $WSe_2$ device

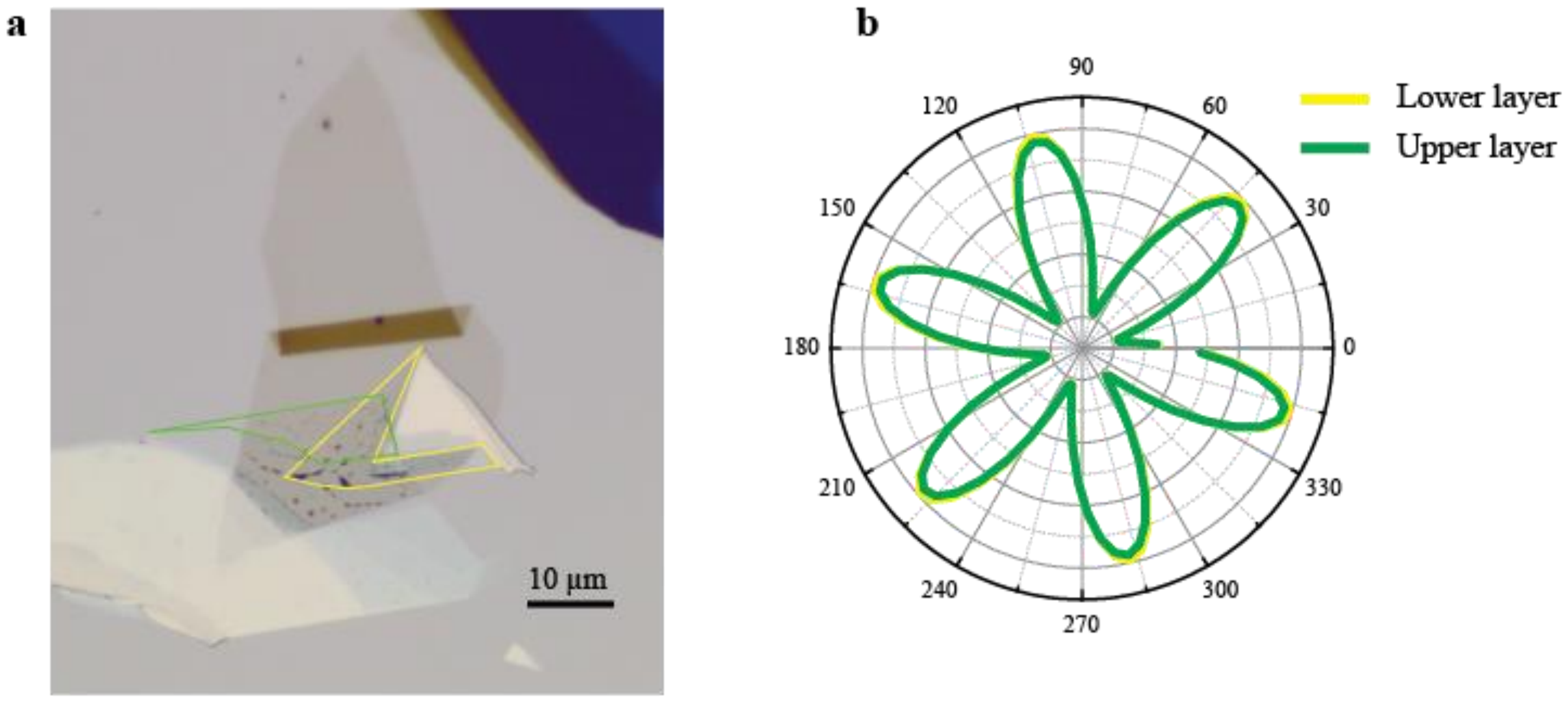

**Figure S1. a,** Optical image of stacked $WSe_2$. Upper and lower monolayers $WSe_2$ are highlighted by green and yellow colours. **b,** Fitted second harmonic generation (SHG) spectroscopy of upper and lower monolayer $WSe_2$ after stacking showing the near zero twist angle.

## 2. AFM topography, PFM amplitude and phase images.

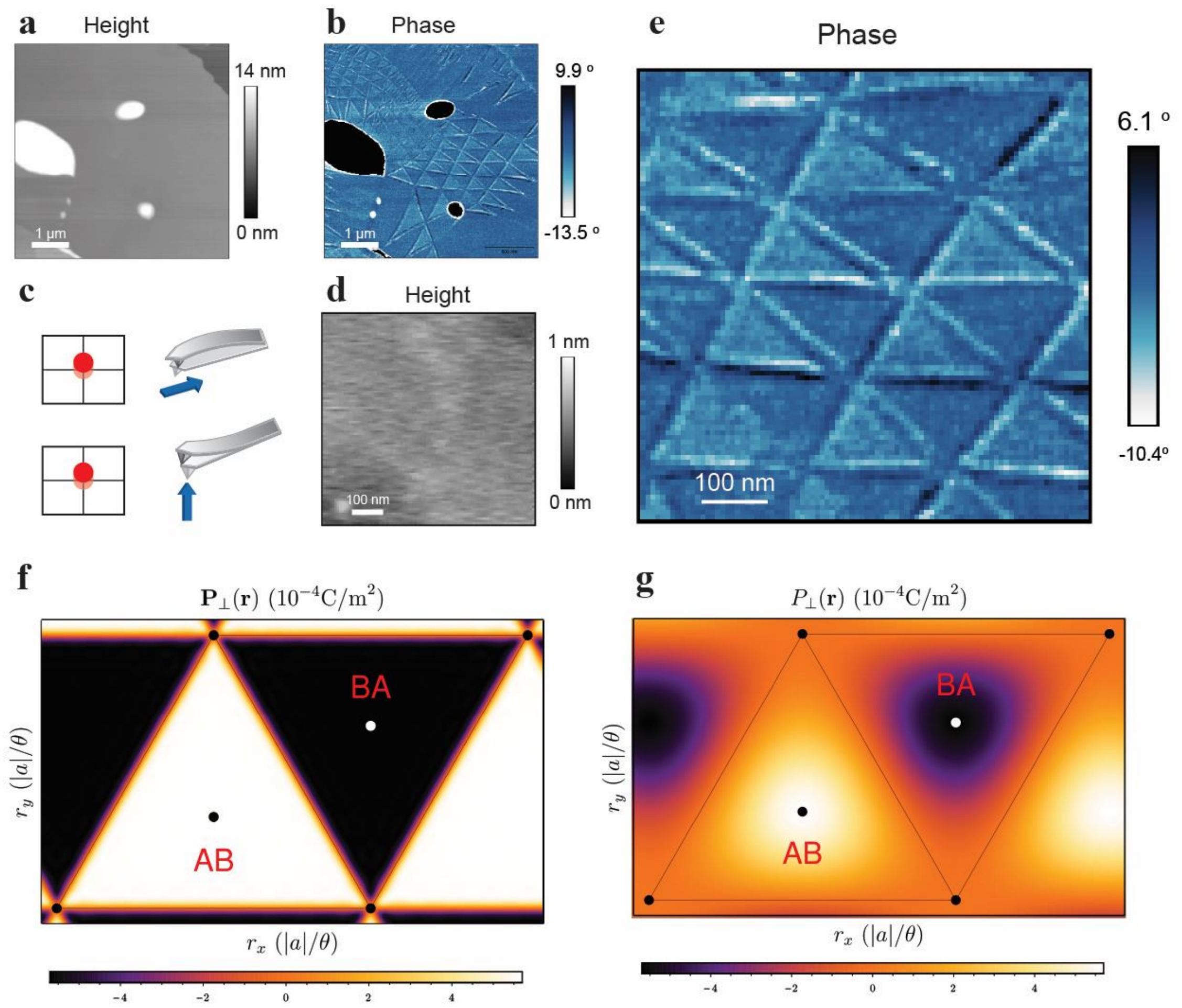

**Figure S2. a, b,** Comparison of topography and phase images obtained of the region shown in Fig. 1(d) of the main manuscript. **c,** Illustration of vertical torsion and deflection resulting from out-of-plane polarization during a vertical PFM measurement. **d, e** Topography and corresponding vertical phase images of a zoomed in region of a,b. There is a strong vertical PFM signal near the DWs because they have large strain gradients which result in a polar response through flexoelectricity[25]. Noted that the in-plane flexoelectric response is negligible[25], see Section S3 for additional details. This couples to piezoelectric stresses, resulting in two strong lines at the DWs in the vertical PFM signals. **f, g,** Out-of-plane polarization in bilayer $WSe_2$ with a twist angle of 0.13°, obtained from **f** DFT calculations and **g** MD calculations. The effects of lattice relaxation on the stacking domains are included.

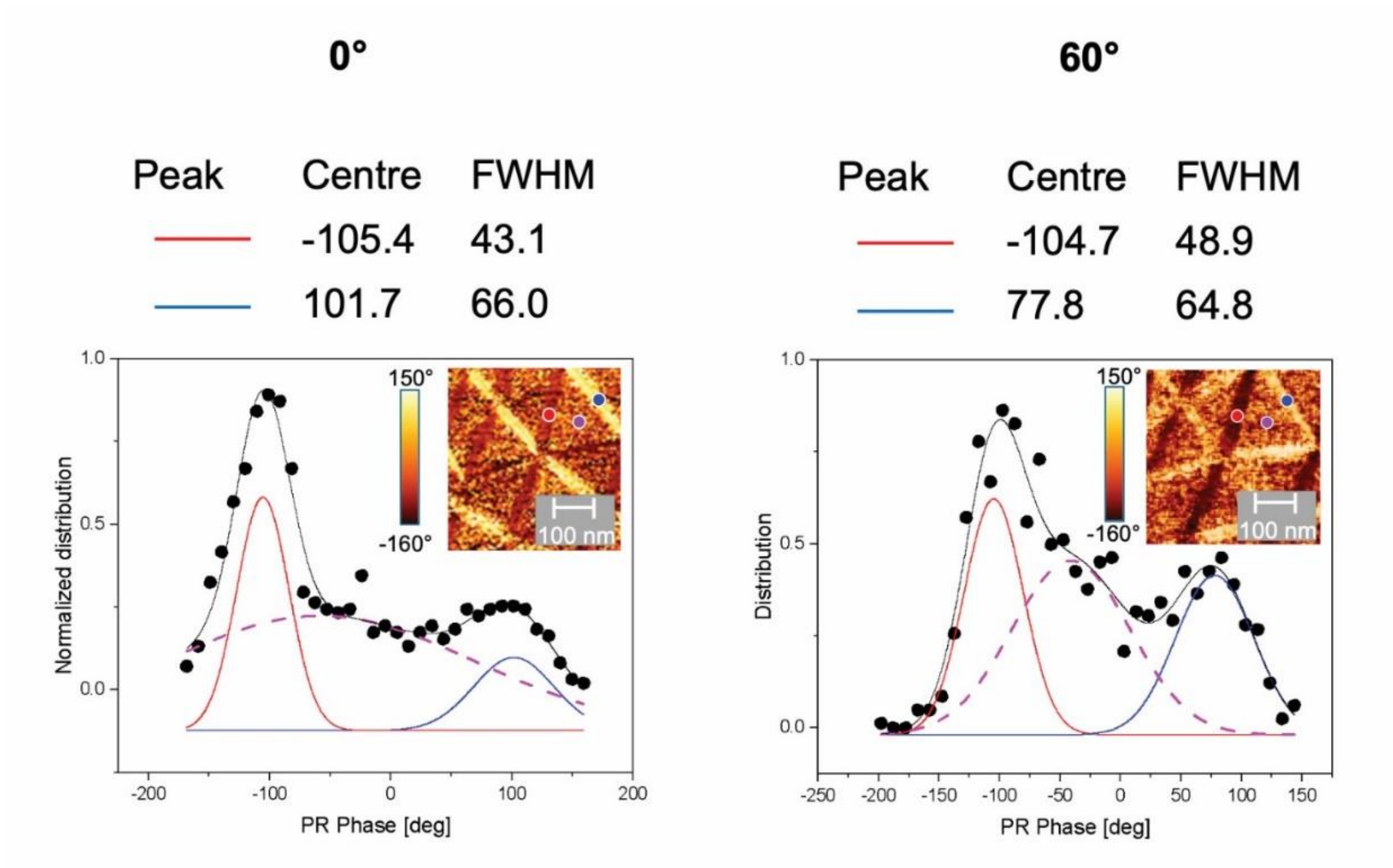

**Figure S3**. Histogram analysis of lateral PFM phase images taken at 0º and 60º showing near 180º phase switch.

### 3. Discussion of in-plane flexoelectricity

The flexoelectric response in twisted bilayers has been considered in detail in McGilly *et al.* [*Nat. Nanotechnol.* **15**, 580–584 (2020)], where they show that the flexoelectric response is localized to the domain walls, where the strain gradient is largest. Furthermore, they show that the most significant flexoelectric response is an out-of-plane polarization arising from a bending mode, with the other (in-plane) responses being negligible.

The in-plane flexoelectric response in twisted bilayers proposed in McGilly *et al.*: Counter-flowing polarization along the outside of the domain walls, with zero polarization exactly in the middle of the domain wall. We do not see this in our measurements: the in-plane polarization flows in the same direction within each direction. Below we compare our measurements with the in-plane polarization arising from flexoelectricity. Therefore, we can rule out in-plane flexoelectricity in our measurements.

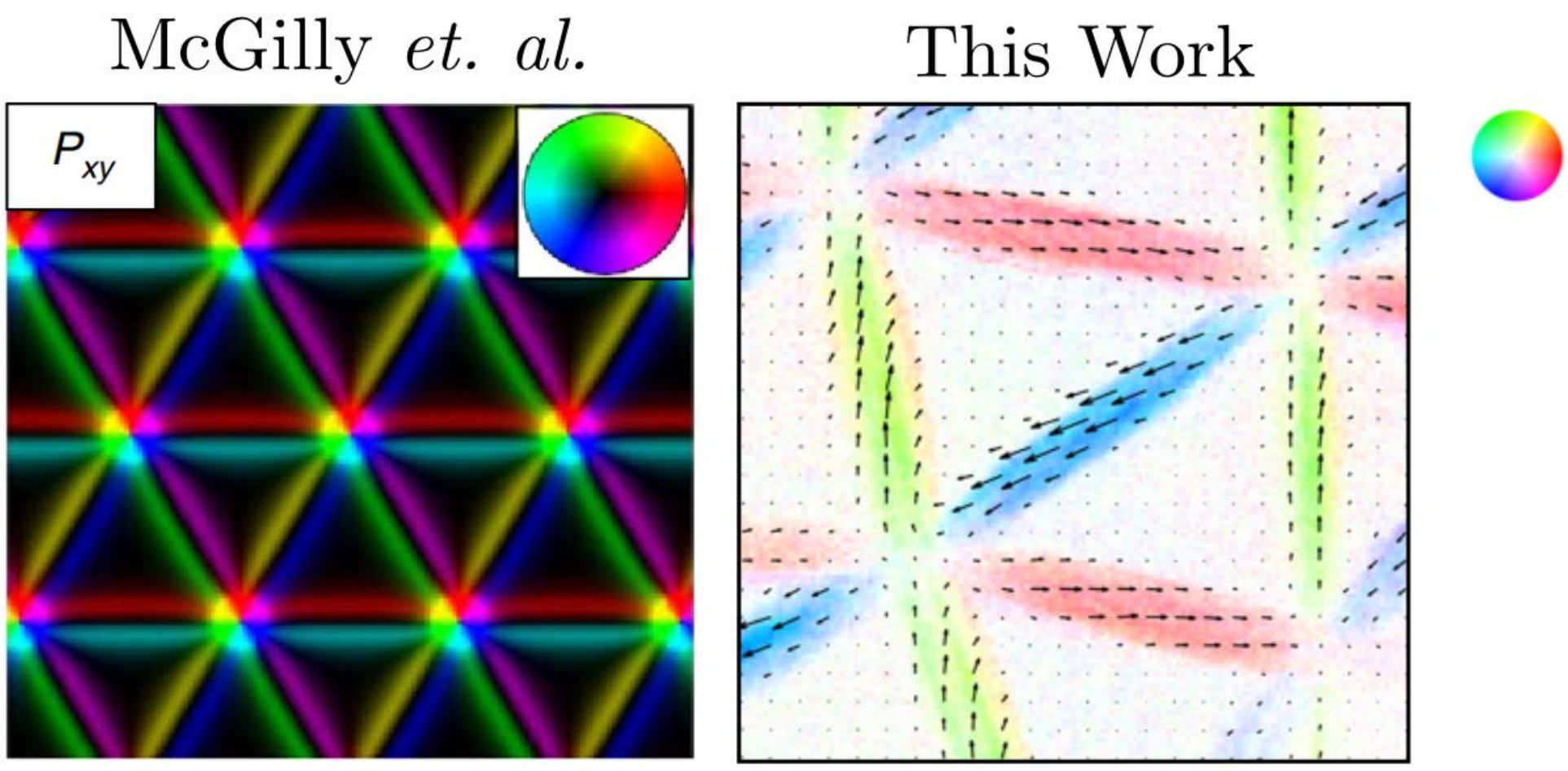

**Figure S4**. Comparison of the in-plane flexoelectric response proposed in McGilly *et al.* (reproduced from [*Nat. Nanotechnol.* **15**, 580–584 (2020)], where only the phase (orientation) is shown, compared to the in-plane polarization observed in our work (right), where both the phase and vector field are shown.

## 4. Generating angle-resolved-PFM (AR-PFM) images

To generate the in-plane angle-resolved-PFM (AR-PFM) images or vector map, the lateral PFM signals taken at different angles (i.e. between sample and cantilever long axis) were determined from their respective amplitude and phase channels. This was achieved by first centring the phase data around 0 degrees and taking the sine of the phase angle multiplied by the amplitude.

Image alignment was conducted using the phase channels as they were clearest. A set of vertices visible in each scan and near to the region of interest were chosen and identified for each scan angle. One scan (the zero-degree angle scan) was designated to be the reference. For the remaining scans, an affine transformation was calculated using the reference vertices, optimised to minimise least squares error.

Once aligned, the x- and y- components of the vector field could be calculated. This was done by projecting the in-phase component along the scan direction for each scan angle. Then the resultant map was created by summing all the contributions and rescaling if necessary.

The resultant vector field was visualised by a hue-saturation map and a superimposed quiver map. For the hue-saturation map, the hue of a pixel corresponds to the angle of the vector and the saturation corresponds to the magnitude of the vector.

## 5. Angular Dependent PFM for 0.03°, 0.04° and 0.05° marginally twisted $WSe_2$ bilayers

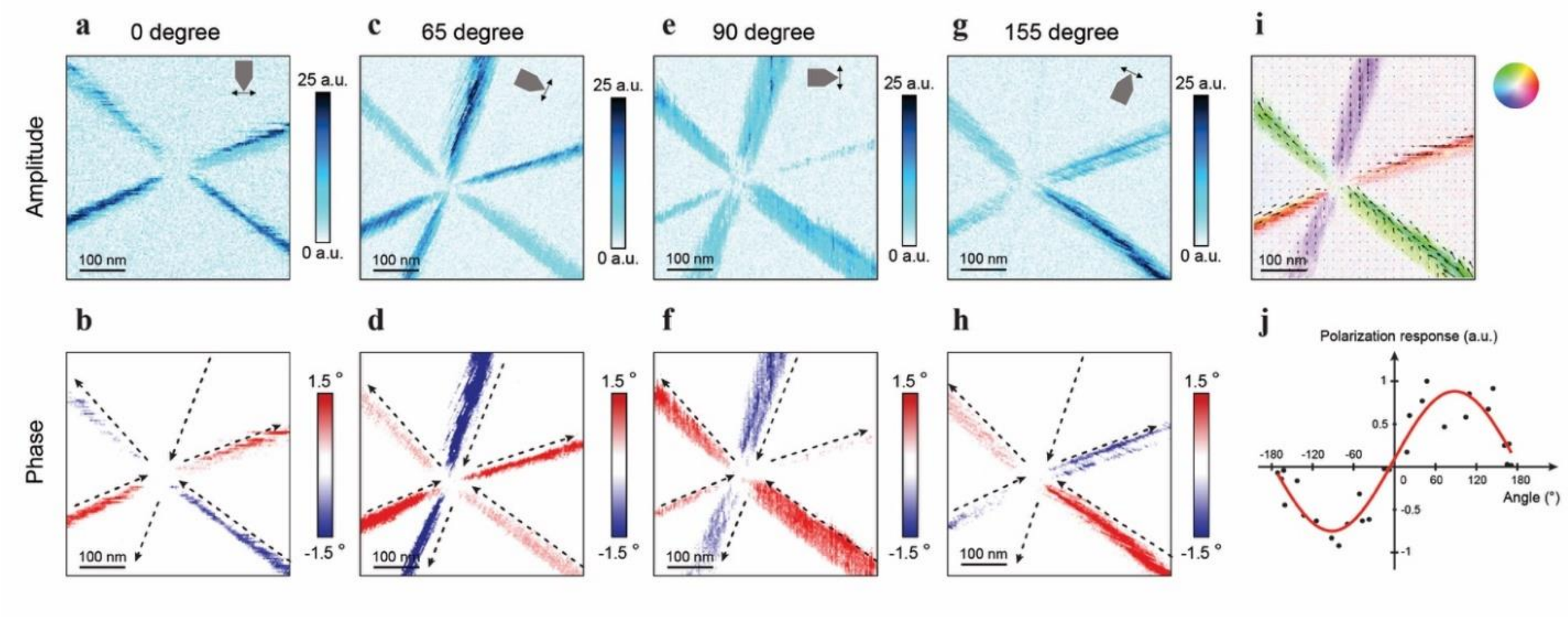

**Figure S5**. Angular dependent PFM of marginally twisted (0.03º) bilayer $WSe_2$**. a-h**, In-plane amplitude and phase images of lateral PFM of the same moiré polar domain when the relative angle between sample and cantilever Φ is **a,b** 0°, **c,d** 65°, **e,f** 90° **g,h** 155°. **i,** Corresponding Angle-Resolved-PFM (AR-PFM) vector map. **j,** Normalized polarization response as a function of in-plane polarization angle φ. The phase offset is -1.7±4.1º.

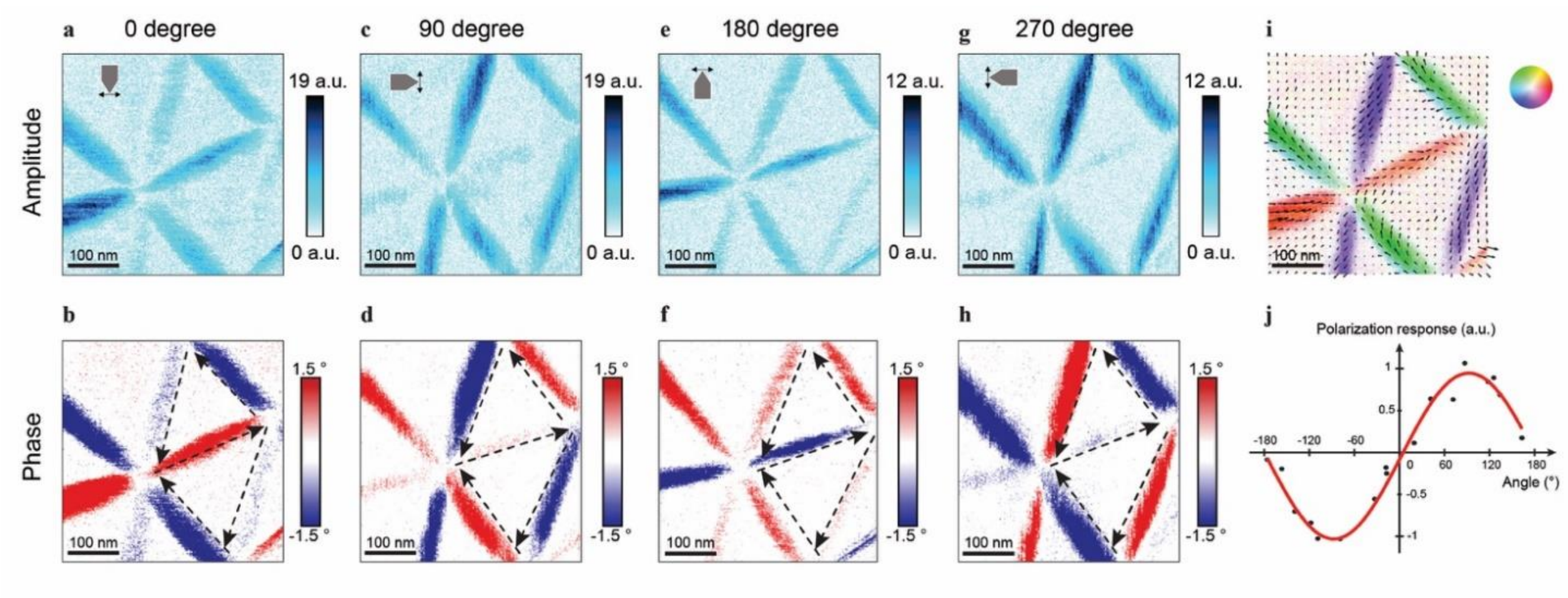

**Figure S6**. Angular dependent PFM of marginally twisted (0.04º) bilayer $WSe_2$**. a-h**, In-plane amplitude and phase images of lateral PFM of the same moiré polar domain when the relative angle between sample and cantilever Φ is **a,b** 0°, **c,d** 90°, **e,f** 180° **g,h** 270°. **i,** Corresponding Angle-Resolved-PFM (AR-PFM) vector map. **j,** Normalized polarization response as a function of in-plane polarization angle φ. The phase offset is 2.4±2.8º.

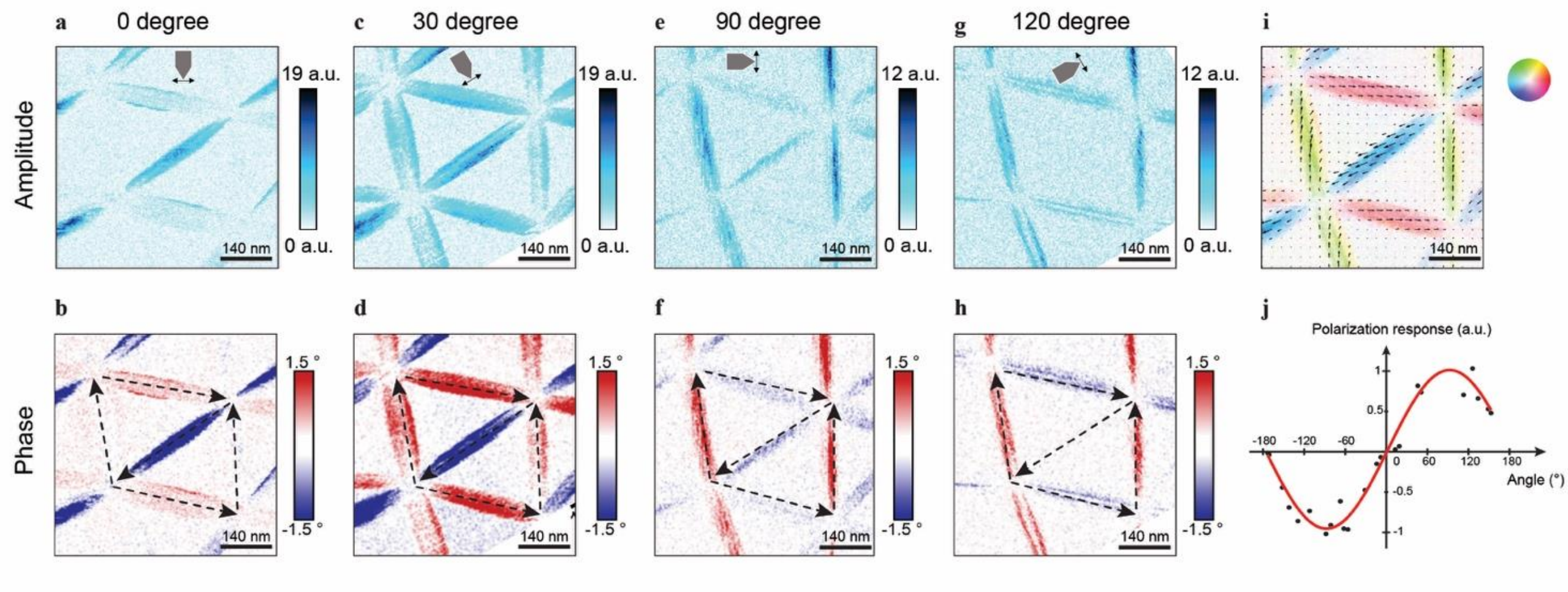

**Figure S7**. Angular dependent PFM of marginally twisted (0.05º) bilayer $WSe_2$**. a-h**, In-plane amplitude and phase images of lateral PFM of the same moiré polar domain when the relative angle between sample and cantilever Φ is **a,b** 0°, **c,d** 30°, **e,f** 90° **g,h** 120°. **i,** Corresponding Angle-Resolved-PFM (AR-PFM) vector map. **j,** Normalized polarization response as a function of the in-plane polarization angle φ. The phase offset is 0.6±2.5º.

## 6. Comparison of 180° and small phase shifts with corresponding vector maps.

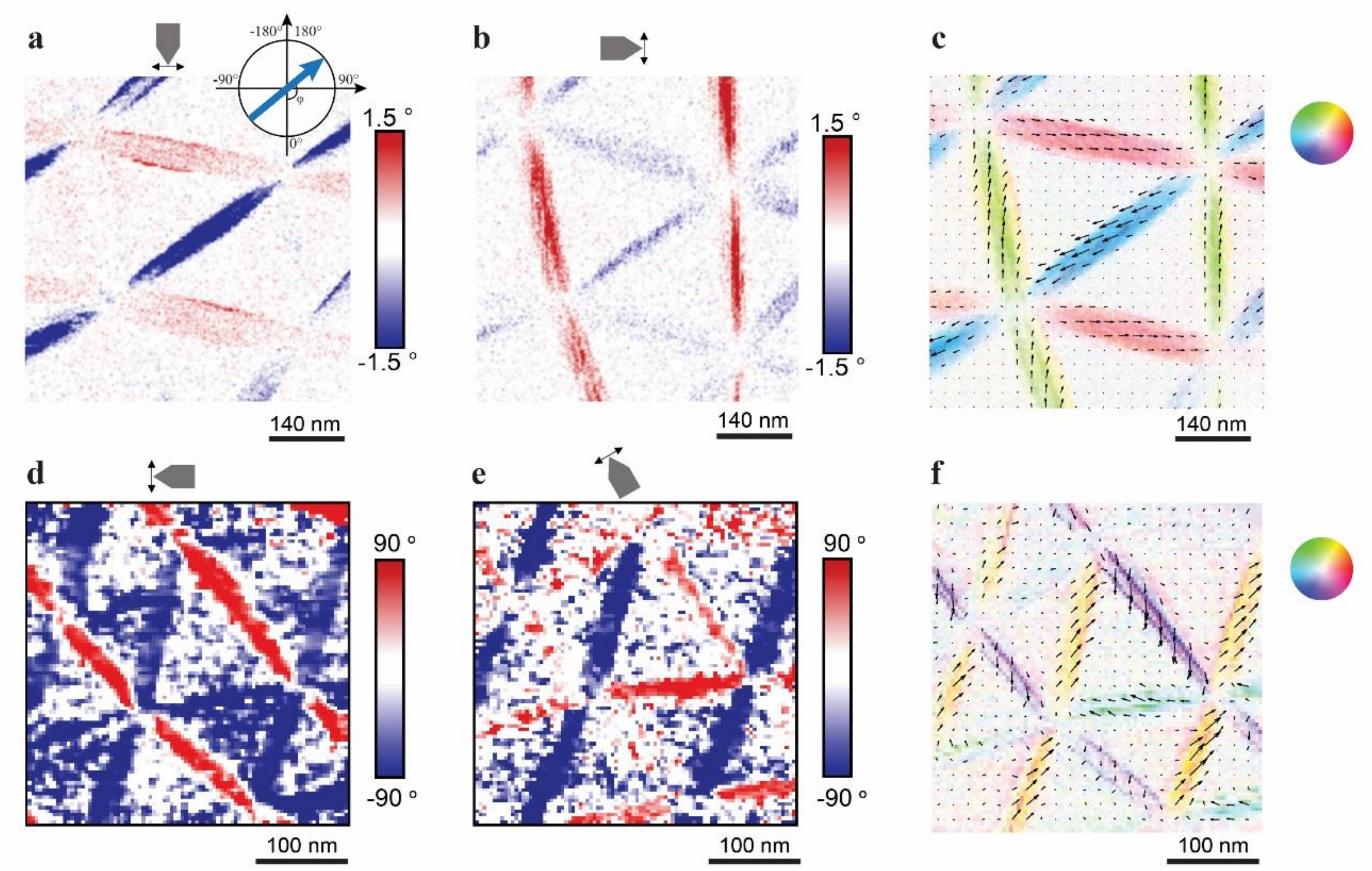

**Figure S8. a-b**, In-plane phase images of lateral PFM of the moiré polar domain when the relative angle between sample and cantilever Φ is **a** 0° **b** 90°. **c,** AR-PFM colorized vector field map generated from lateral PFM images in **a**–**b**. **d-e**, In-plane phase images of lateral PFM of the moiré polar domain when the relative angle between sample and cantilever Φ is **d** 0° **e** 60°. **f,** AR-PFM colorized vector field map generated from lateral PFM images in d–**e.**

## 7. PFM contact resonance frequency curve

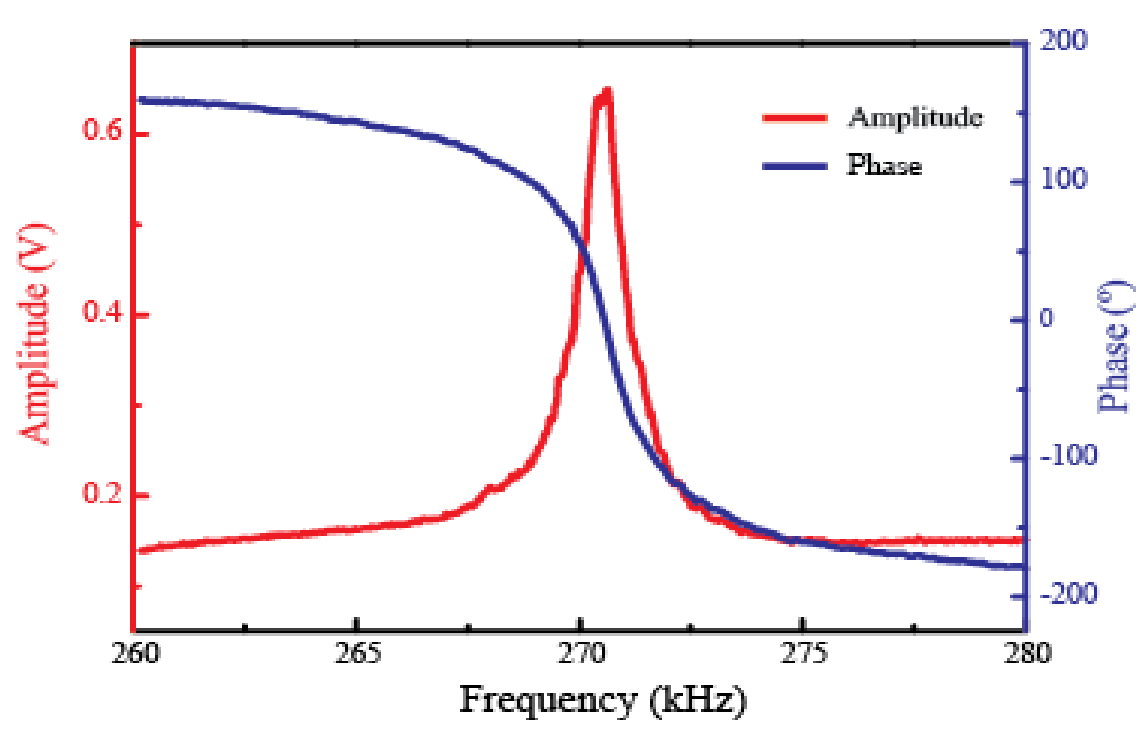

**Figure S9.** Contact resonance frequency in PFM.

## 8. Comparing measurement near-resonance and off-resonance frequencies

To achieve more stable signals during repeated piezoresponse mapping and to minimize cross-talk between in-plane and out-of-plane responses – issues that are amplified in resonance-enhanced modes such as dual AC resonance tracking – we decided to use off-resonance conditions. Importantly, the results obtained at both near-resonance and off-resonance frequencies showed no significant differences in the measured domain structures as shown in Fig. S10 below. All features across the three samples remain the same, and consistent with winding of the polarization.

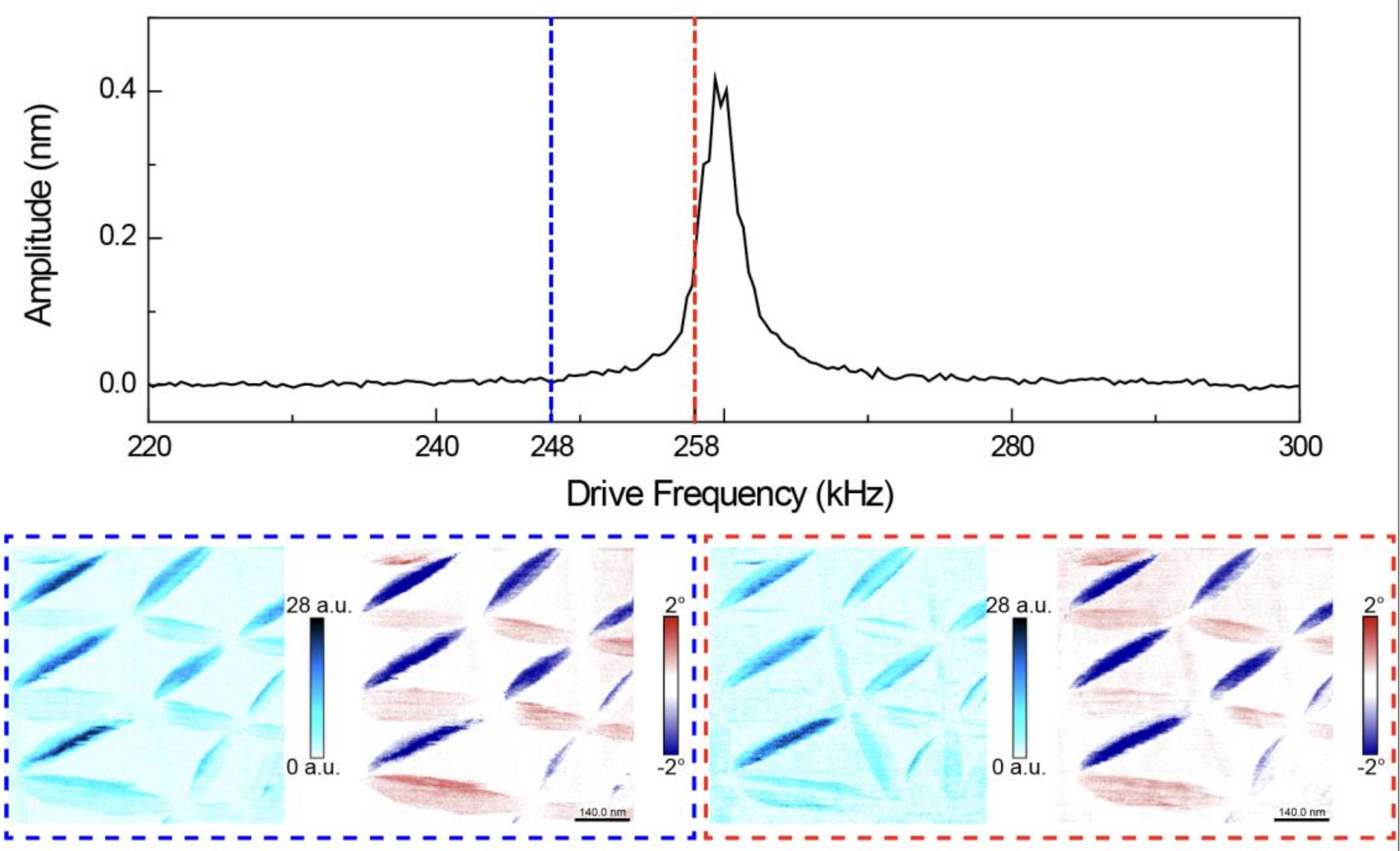

**Figure S10**. The top graph shows the contact resonance frequency spectrum in PFM, where the near-resonance and off-resonance drive frequencies are marked with red and blue dashed lines, respectively. The bottom images display the corresponding PFM measurements. Images in the red box represent near-resonance lateral amplitude and phase, images in the blue box show off-resonance lateral amplitude and phase.

## 9. Measurements with tip radius of 5 nm and 25 nm

We performed additional measurements using a PFM tip with a significantly smaller tip radius: 5 nm compared with the regular tip which has a 25 nm radius. After performing measurements with higher resolution, we observed a narrowing of the domain walls, but the winding of the polarization was consistently observed.

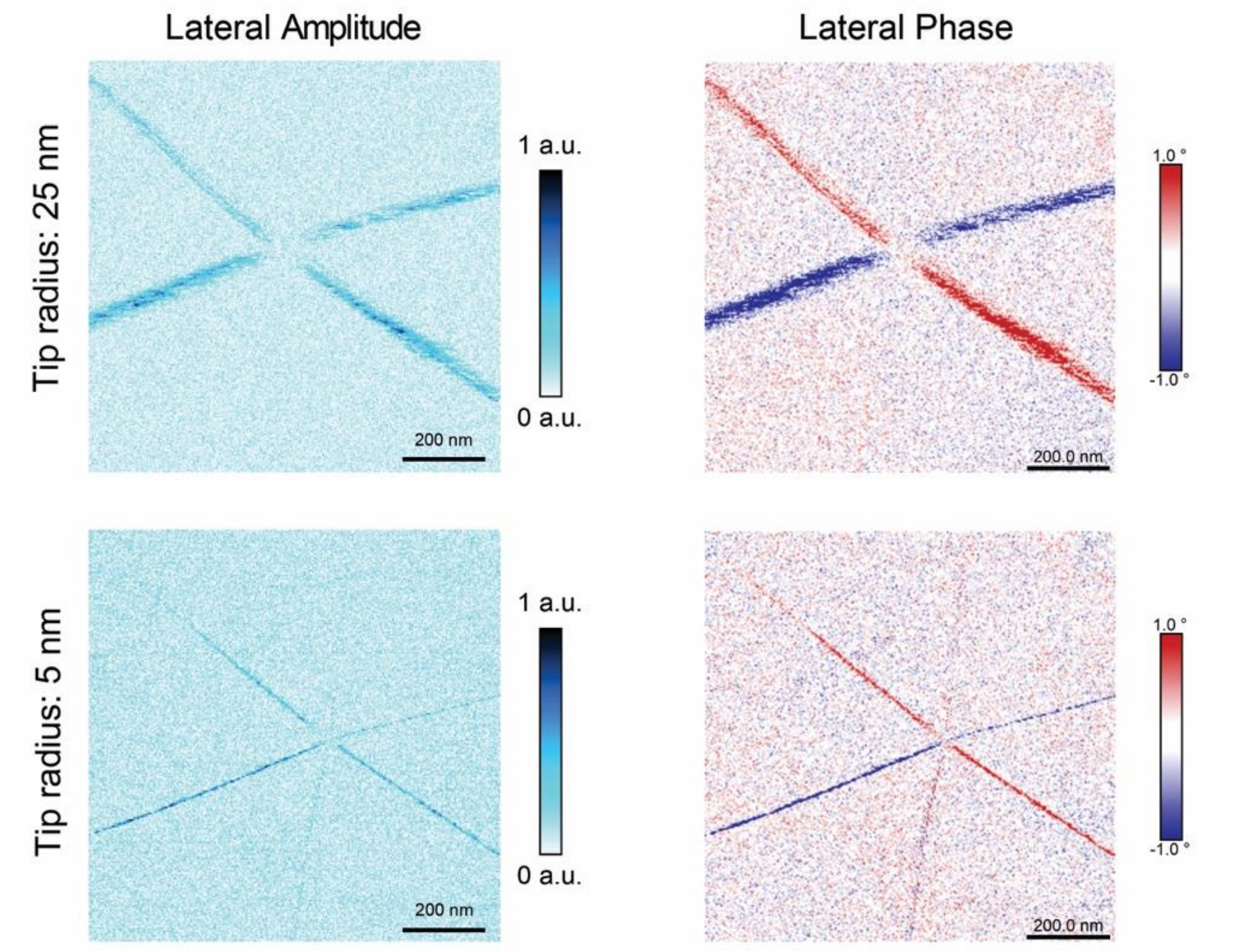

**Figure S11**. Lateral PFM amplitude and phase images were acquired for the same region using a tip with a 5 nm and 25 nm radius.

**10. Molecular dynamics simulations of temperature dependence of twisted $WSe_2$ bilayers.**

To investigate the temperature dependence of the winding, we performed molecular dynamics simulations to investigate the stability of twisted $WSe_2$ bilayers. The $WSe_2$ bilayers were first relaxed using standard minimization. Once the structure was relaxed, molecular dynamics simulations were performed in the NVT ensemble (constant number of particles, volume, and temperature). The simulations were performed at two distinct temperatures: 100 K and 300 K. Fig. S12 shows the fluctuations of the average interlayer separation as a function of time. The results show that in both temperature regimes, the system equilibrates over time, with the interlayer distance fluctuating around a stable equilibrium value. At both temperatures, the system initially exhibits significant fluctuations that diminish as the system approaches equilibrium. This behaviour is characteristic of systems transitioning from a non-equilibrium to an equilibrium state, where thermal energy is dissipated, and the system stabilizes at its minimum energy configuration. These results clearly demonstrate that at finite temperatures, the $WSe_2$ bilayer undergoes thermal fluctuations around an equilibrium interlayer distance. The extent of these fluctuations appears to reduce as the system equilibrates, suggesting that the layers are dynamically stable, with the interlayer distance consistently fluctuating around a stable value after the equilibrium is reached. This analysis offers valuable insights into the thermal stability of $WSe_2$ layers, indicating that they remain in a stable configuration at both 100 K and 300 K, despite the inherent thermal fluctuations at higher temperatures.

The shape of the polarization field depends on the local stackings in the twisted supercell. Because the stackings do not dramatically change at room temperature, only fluctuating slightly about equilibrium values, the winding of the polarization and its topological nature persists at room temperature. This is further supported by our measurements of the winding, performed at room temperature, which are in excellent agreement with theoretical predictions (zero temperature).

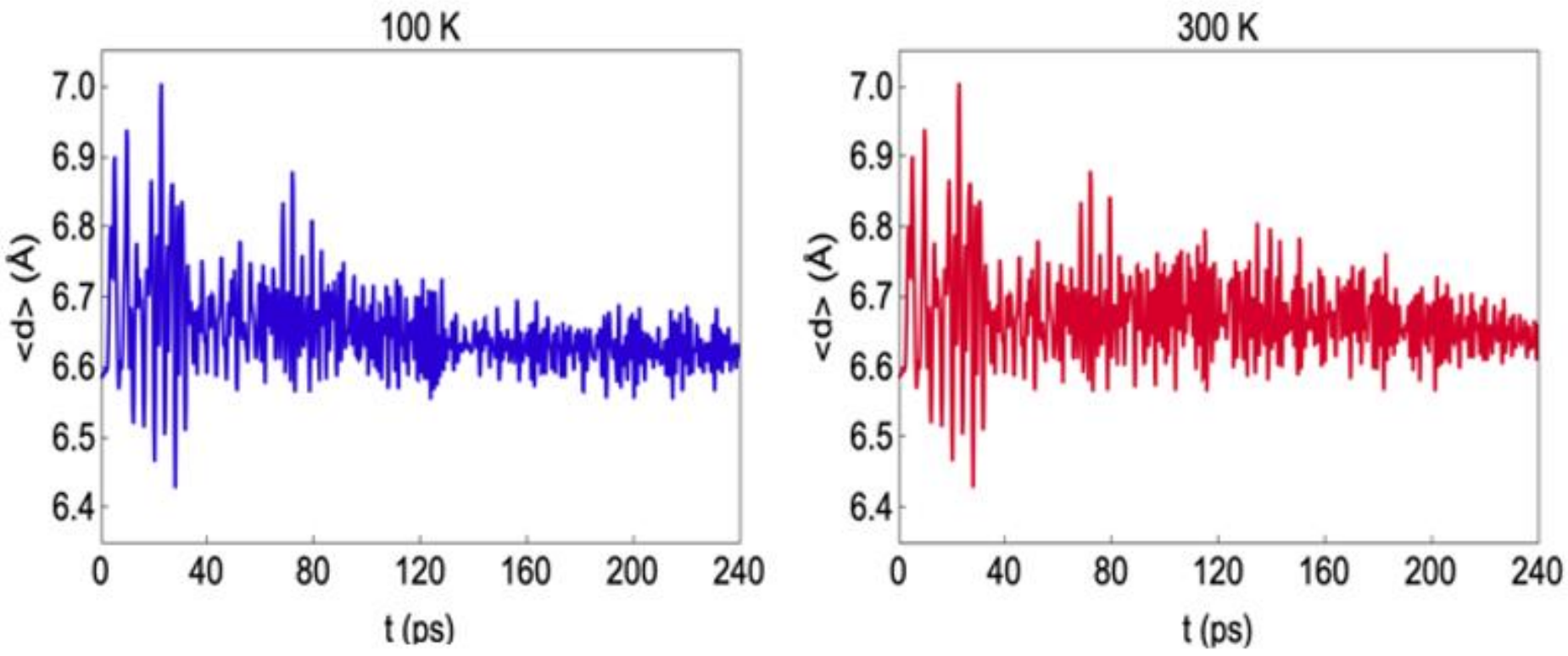

**Figure S12.** Fluctuations in the average interlayer distance <d> as a function of time t for temperatures T=100 K and T=300 K as obtained from LAMMPS molecular dynamics simulations. The plot shows the temporal evolution of <d> at each temperature, highlighting the differences in thermal motion and system behaviour at lower and higher temperatures.

**11. Periodically-poled lithium niobite standard sample measurement at off-resonance frequency**

To verify there was no cross-talk from out-of-plane polarization to lateral PFM, we performed measurements on the Bruker standard PFM sample, which consists of periodically poled lithium niobate with pure out-of-plane polarization. Using the same off-resonance frequency settings as for our twisted sample, we observed a 180° phase difference in vertical PFM, consistent with out-of-plane polarization switching. Meanwhile, the lateral PFM showed minimal signal, indicating negligible contribution of the vertical domains to the lateral signal at domain walls under our experimental setup. This minimal signal is significantly smaller than the lateral PFM signals observed in our twisted bilayers.

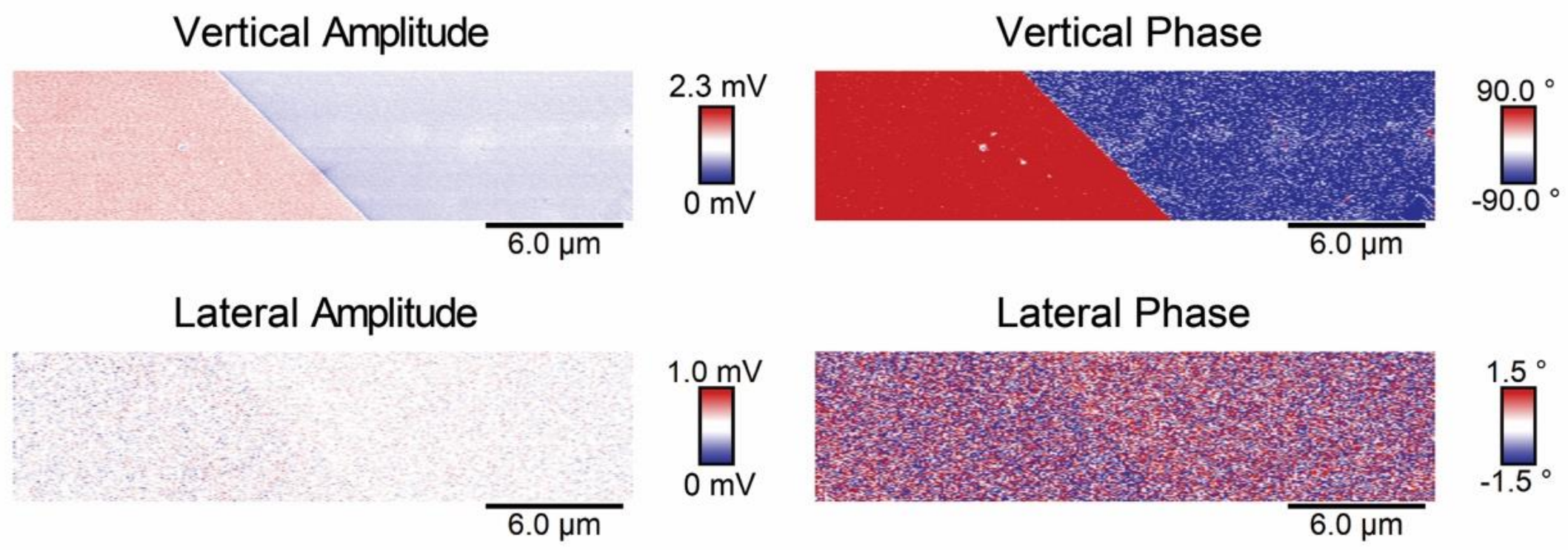

**Figure S13**. PFM images of periodically-poled lithium niobite standard sample.